\documentstyle[12pt,colordvi,epsf,epsfig]{article}
\newfont{\thiplo}{msbm10 scaled\magstep 2}
\newfont{\gothic}{eufb10 scaled\magstep 2}
\newfont{\unc}{eurb10} 
\newskip\humongous \humongous=0pt plus 1000pt minus 1000pt
\def\caja{\mathsurround=0pt}\def\eqalign#1{\,\vcenter{\openup1\jot \caja
        \ialign{\strut \hfil$\displaystyle{##}$&$ 
        \displaystyle{{}##}$\hfil\crcr#1\crcr}}\,}
\newif\ifdtup

\def\eqright #1\cr{\noalign{\hfill$\displaystyle{{}#1}$}}
\def\eqleft #1\cr{\noalign{\noindent$\displaystyle{{}#1}$\hfill}}

\def\oldreffmt#1{\rlap{[#1]} \hbox to 2\parindent{}}

\def\figfmt#1{\rlap{Figure {#1}} \hbox to 1in{}}

\def\sectioneq{\def\theequation{\thesection.\arabic{equation}}{\let
\holdsection=\section\def\section{\setcounter{equation}{0}\holdsection}}}%

\newcounter{holdequation}

\def\begineq #1\endeq{$$ \refstepcounter{equation}\eqalign{#1}\eqno
	(\theequation) $$}
\def\contlimit{\,{\hbox{$\longrightarrow$}\kern-1.8em\lower1ex
\hbox{${\scriptstyle (a\rightarrow0)}$}}\,}
\def\centeron#1#2{{\setbox0=\hbox{#1}\setbox1=\hbox{#2}\ifdim
\wd1>\wd0\kern.5\wd1\kern-.5\wd0\fi
\copy0\kern-.5\wd0\kern-.5\wd1\copy1\ifdim\wd0>\wd1
\kern.5\wd0\kern-.5\wd1\fi}}
\def\centerover#1#2{\centeron{#1}{\setbox0=\hbox{#1}\setbox
1=\hbox{#2}\raise\ht0\hbox{\raise\dp1\hbox{\copy1}}}}
\def\centerunder#1#2{\centeron{#1}{\setbox0=\hbox{#1}\setbox
1=\hbox{#2}\lower\dp0\hbox{\lower\ht1\hbox{\copy1}}}}
\def\lsim{\;\centeron{\raise.35ex\hbox{$<$}}{\lower.65ex\hbox
{$\sim$}}\;}
\def\gsim{\;\centeron{\raise.35ex\hbox{$>$}}{\lower.65ex\hbox
{$\sim$}}\;}

\def\super#1{\ifmmode \hbox{\textsuper{#1}}\else\textsuper{#1}\fi}
\def\textsuper#1{\newcount\holdspacefactor\holdspacefactor=\spacefactor
$^{#1}$\spacefactor=\holdspacefactor}

\def\getcite#1,{\advance\citenumber by1
\def\getcitearg{#1}\def\lastarg{@}
\ifnum\citenumber=1
\ref{#1}\let\next=\getcite\else\ifx\getcitearg\lastarg\let\next=\relax
\else ,\ref{#1}\let\next=\getcite\fi\fi\next}

\def\pom{{\rm P\kern -0.53em\llap I\,}}
\def\spom{{\rm P\kern -0.36em\llap \small I\,}}
\def\sspom{{\rm P\kern -0.33em\llap \footnotesize I\,}}

\relax
\def\contlimit{\,{\hbox{$\longrightarrow$}\kern-1.8em\lower1ex
\hbox{${\scriptstyle (a\rightarrow0)}$}}\,}
\def\upon #1/#2 {{\textstyle{#1\over #2}}}
\relax
\renewcommand{\thefootnote}{\fnsymbol{footnote}}

\sectioneq
\def\subhead#1{\bigskip\vbox{\noindent\bf #1}\nobreak\par}

\def\til#1{\centeron{\hbox{$#1$}}{\lower 2ex\hbox{$\char'176$}}}
\def\tild#1{\centeron{\hbox{$\,#1$}}{\lower 2.5ex\hbox{$\char'176$}}}
\def\sumtil{\centeron{\hbox{$\displaystyle\sum$}}{\lower
-1.5ex\hbox{$\widetilde{\phantom{xx}}$}}}

\begin{document}
\begin{titlepage} 

\rightline{\vbox{\halign{&#\hfil\cr
&\today\cr}}} 
\vspace{0.25in} 

\begin{center} 

Past, Present, and Future Multi-Regge Theory

\medskip

Alan R. White\footnote{arw@anl.gov }

\vskip 0.6cm

\centerline{Argonne National Laboratory}
\centerline{9700 South Cass, Il 60439, USA.}
\vspace{0.5cm}

\end{center}

\begin{abstract} 

 The connection of the unitary Critical Pomeron to QUD - a unique massless, infra-red fixed-point, left-handed SU(5) field theory that might provide an unconventional underlying unification for the Standard Model, is discussed in the context of
 developments in past, present, and future multi-regge theory.  
 The QUD bound-state S-Matrix is accessible via elaborate (non-planar) multi-regge theory. Standard Model states and interactions are replicated via massless fermion anomaly dynamics in which configurations of infra-red divergent anomalous gauge boson reggeons play a wee parton vacuum-like role. All particles, including neutrinos, are bound-states with dynamical masses and there is no Higgs field. 
 A color sextet quark sector, that could be discovered at the LHC, produces both Dark Matter and Electroweak Symmetry Breaking and the very small QUD coupling should be reflected in the smallness of neutrino masses.
 The origin of the Standard Model could be 
that it is reproducing the unique, unitary, S-Matrix!

\end{abstract} 

\vspace{0.5in}

\begin{center}

{\it Presented at ``Scattering Amplitudes and the Multi-Regge Limit'', 
\newline Feb 10-14, 2014, Madrid, Spain.}
\end{center}

\renewcommand{\thefootnote}{\arabic{footnote}} 

\end{titlepage} 

Although the title of this talk suggests a general review, the main focus will be on very
exciting results, involving the potential origin of the Standard Model, that I have seen gradually emerge from my work in the last few years. It is very unfortunate that 
the multi-regge framework within which I  work is unfamiliar to most physicists. However, it is very close to the topic of this workshop  and since the participants are correspondingly specialized, I am hopeful that (at least some of) my argumentation will be appreciated. The main novelty in my work, compared to more mainstream multi-regge theory, is the utilisation of general non-planar
multi-regge kinematics and corresponding gauge theory reggeon diagrams to obtain bound-state amplitudes formed via infra-red divergences coupled to massless fermion anomalies.

$~$
  
\noindent OUTLINE
\begin{enumerate}
\item{{\bf Abstract S-Matrix Multi-Regge Theory:} 
reggeon unitarity, multi-regge limits, asymptotic dispersion theory and multi-regge amplitudes.}
\item{{\bf The Regge Pole Critical Pomeron and QCD:} experiment motivates the formulation of the Critical Pomeron. Basic properties of the supercritical Pomeron suggest it
occurs in Superconducting QCD.}
\item{{\bf Multi-Regge Gauge Theory:} brief comments on 2-to-N, non-planar N-to-N; including massless fermion anomaly vertices and anomaly poles, infra-red reggeon kernels, wee-partons in superconducting QCD bound-state amplitudes.}
\item{{\bf The Critical Pomeron in Massless QCD$_S$:} an experimentally desirable pomeron and confinement spectrum. The chirality transition critical phenomenon. }
\item{{\bf The Electroweak Interaction plus Massless QCD$_S$:} the selection of QUD - a unique massless SU(5) theory. Infra-red reggeon diagram construction of the multi-regge S-Matrix gives the  Critical Pomeron and the Standard Model? The Standard Model is reproducing the Unique Unitary S-Matrix?? Electroweak symmetry breaking, three generations, dark matter, neutrino masses, all emerge.}
\item{{\bf The Nightmare Scenario:} ``We Got it Wrong. How did we misread the signals? What to Do?''}
\item{{\bf The Future:} At the LHC, multi-regge theory (with very extensive development) could 
be essential for study of the large cross-section high-mass sextet quark sector of the QUD S-Matrix.}
\end{enumerate}

\newpage 

\subhead{1. Abstract S-Matrix Multi-Regge Theory}

\vspace{0.1in}
In effect, the ``birth'' of multi-regge theory was 50 years ago - a brilliant, seminal, paper by Gribov, Pomeranchuk and Ter-Martirosyan\cite{gpt}. It was shown how multiparticle
$t$-channel unitarity in the j-plane produces ``reggeon unitarity'' discontinuities  for regge cuts. Ambiguities in the analysis led Gribov to introduce (via model field theories) RFT reggeon diagrams that explicitly satisfy reggeon unitarity,.
 
A fundamental derivation of reggeon unitarity requires dispersion relation based multiparticle regge theory. However, for a long time it was feared that because the analyticity properties of multiparticle amplitudes are incredibly complicated, such a derivation would not be 
possible. Very fortunately, it was discovered\cite{arw1,hps,arw2,sw,arw3} that in general multi-regge (multi-rapidity) limits, defined via 3-point vertex tree diagram angular variables, the analytic structure simplifies enormously. As a result, asymptotic dispersion relations can be written that give a general multiparticle amplitude as a sum of
``hexagraph'' amplitudes, each of which has distinct asymptotic cuts satisfying the Steinmann relations. It can then be shown that hexagraph amplitudes have corresponding angular momentum and helicity plane transforms that are sufficient\cite{arw1,arw3} to develop full multiparticle complex angular momentum theory. 

As an example, a simple tree diagram is shown in Fig.~1(a), with one correponding hexagraph shown in
Fig.~1(b). 
\vspace{0.1in}
\begin{center}
\epsfxsize=2.5in\epsffile{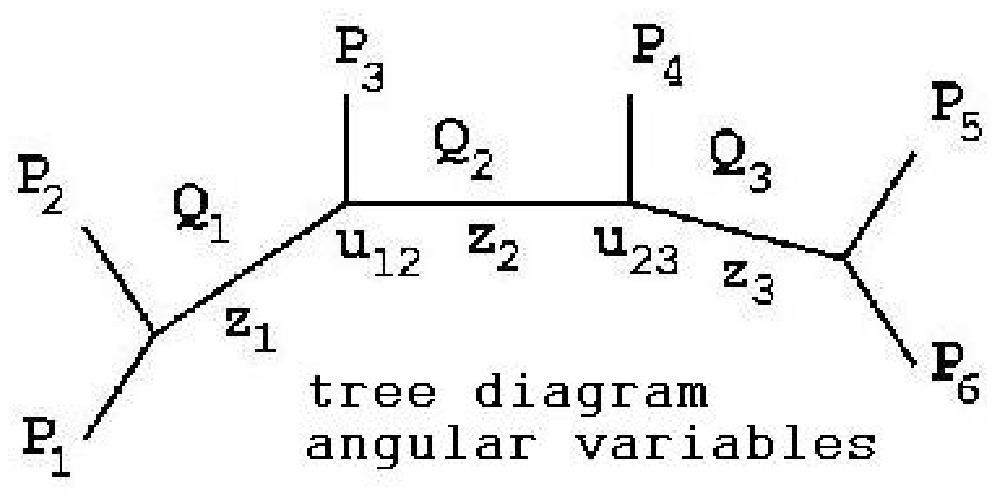} \hspace{0.2in} \epsfxsize=2.5in\epsffile{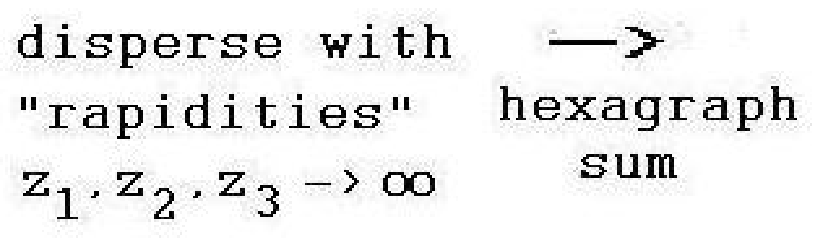} 

(a) 

\epsfxsize=5.5in\epsffile{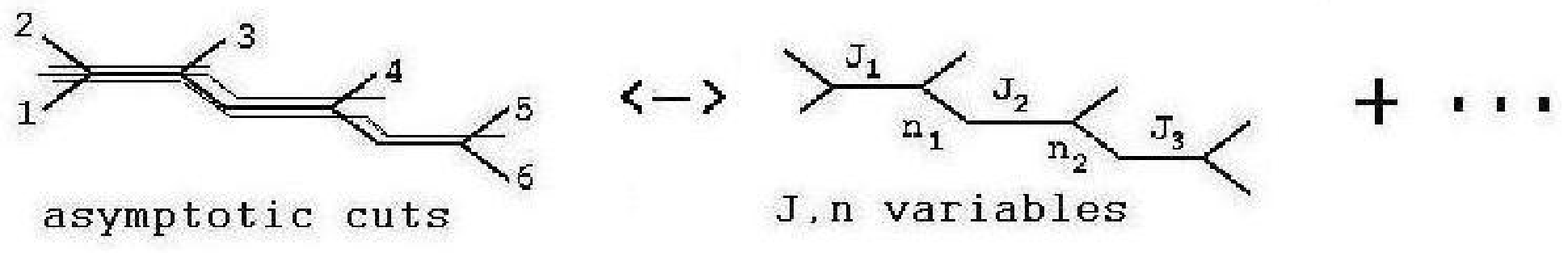}

(b)

\vspace{0.1in}
Figure 1 (a) A Simple Tree Diagram (b) A Corresponding Hexagraph
\end{center}

The partial-wave expansion corresponding to the tree diagram is
$$
\eqalign{
M_N(\til{t},g_1,\ldots,g_{N-3})={}&\sum^\infty_{J_1=0}\,
\sum_{|n_1|,|n'_1|<J_1}\ldots\sum^\infty_{J_{N-3}=0}~~\,
\sum^\infty_{|n_{N-3}|,|n'_{N-3}|<J_{N-3}}\cr
&D^{J_1}_{n_1n'_1}(g_1)\ldots D^{J_{N-3}}_{n_{N-3},n'_{N-3}}(g_{N-3})
a_{J_1,n_1,n'_1,\ldots,J_{N-3},n_{N-3},n'_{N-3}}(\til{t}).}
$$
The Sommerfeld-Watson representation corresponding to the hexagraph of Fig.~1(b) is
$$
\eqalign{                                  
A_H={}{1\over 8}\int_{C_{n_2}}
{dn_2\,u_2^{n_2}\over \sin\pi
n_2}~&\int_{C_{n_1}} {dn_1\,u_1^{n_1}\over
\sin\pi(n_1-n_2)}~\int_{C_{J_1}} {dJ_1~d^{J_1}_{0,n_1}(z_1)\over
\sin\pi(J_1-n_1)}\cr
&\times~   \sum^\infty_{{\scriptstyle J_2-n_1=N_1=0\atop
\scriptstyle J_3-n_2=N_2=0}}
d^{J_2}_{n_1,n_2}(z_2)d^{J_3}_{n_2,0}(z_3)a_{N_2N_3}(J_1,n_1,n_2,
\til{t}) ~~ + ~~\cdots
,}
$$
In the multi-regge limit, corresponding to the tree diagram, regge poles at $ j_i = \alpha_i(t_1), i=1,2,3$ give 
$$
\eqalign{A_H  ~  \centerunder{\centerunder{ 
\centerunder{$ \sim$}{\raisebox{-2mm}{$\scriptstyle
z_1\to\infty $}}}{\raisebox{-2mm}{$\scriptstyle z_2\to\infty$}}}{
\raisebox{-2mm}{$\scriptstyle z_3\to\infty $}}~
&{z_1^{\alpha_1}z_2^{\alpha_2}z_3^{\alpha_3}\over 
\sin\pi\alpha_3\sin\pi(\alpha_2-\alpha_3)\sin\pi(\alpha_1-\alpha_2)}
\sum^\infty_{N_1=N_2=0}\biggl[\beta_{N_1N_2}^{\alpha_1\alpha_2\alpha_3}
u_1^{\alpha_2-N_1}u_2^{\alpha_3-N_2} \cr
&~~~~~~~~~~+ \beta_{N_1N_2}^{\alpha_1 -\alpha_2\alpha_3}
u_1^{- \alpha_2-N_1}u_2^{\alpha_3-N_2} + 
\beta_{N_1N_2}^{\alpha_1\alpha_2-\alpha_3} 
u_1^{\alpha_2-N_1}u_2^{-\alpha_3-N_2}\cr
&~~~~~~~~~~~~~~~~~~~~~~ + 
\beta_{N_1N_2}^{\alpha_1-\alpha_2-\alpha_3}
u_1^{-\alpha_2-N_1}u_2^{-\alpha_3-N_2}\biggr]. }$$

The most important result, for our present purposes, is that reggeon unitarity generalizes to  all channels of multi-regge amplitudes,
implying that gauge theory reggeon diagrams derived in elementary
amplitudes, appear analagously in multi-regge amplitudes.
It is very important that regge poles and related regge cuts are the only 
j-plane singularities that are consistent with reggeon unitarity. Eikonal amplitudes are not !!


\subhead{2. The Regge Pole Critical Pomeron and QCD:}

$~$ 

It is now unambiguously clear that the experimental
forward region pomeron is, in first approximation,  an even signature regge pole
\footnote{Experimentally, in hadron scattering, the BFKL pomeron is, at most, present at large transverse momentum.}. Data from all hadron forward scattering experiments\cite{bu,arw4}, up to and, including the ISR experiments show the same, universal, shrinking, diffraction peak - with factorizing cross-sections. In recent years, unfortunately, the neglect of the forward region has made it difficult to reinforce this statement at higher energies. Remarkably, however, the TOTEM experiment has very recently shown that the universal low-energy shrinkage, extrapolated to higher energy, is exactly what is seen at the LHC\cite{tot}.

Motivated by the experimental situation, the Critical Pomeron was formulated\cite{cri} as an even signature regge pole with a triple pomeron vertex $ir$, and with intercept $~\alpha_{\spom}(0) \to 1$.  All the
multi-pomeron cuts accumulate when $~\alpha_{\spom}(0) \to 1$, but an RFT fixed-point solution exists. The $\epsilon$ - expansion shows (in analogy with similar applications) that a fixed-point appears in the $\beta$-function 
as the transverse momomentum dimension is lowered from the scaling value of four,  towards the physical value of two. This expansion can also be used to consistently calculate full diffractive amplitudes. Crucially, the $\epsilon$ expansion does not imply, in any way, that the fixed-point theory only exists near $\epsilon = 0$. Indeed, all the evidence is to the contrary.

In effect, the Critical Pomeron is the summit of S-Matrix Theory\cite{arw3}. It, uniquely, satisfies all known unitarity constraints on asymptotically rising cross-sections. Rising cross-sections are, presumably, essential if an underlying asymptotically free gauge theory is involved?? 

As we elaborate on in the following, it is the ``supercritical pomeron'', defined with the bare pomeron intercept $\alpha^0_{\spom} (0) > 1$, that actually provides a path to an underlying field theory.
Because of the non-hermiticity of the triple pomeron interaction, the formulation of the supercritical theory is non-trivial,  There is no (conventional) supercritical ``pomeron condensate'' expansion satisfying reggeon unitarity. Instead a pomeron condensate has to be introduced\cite{arw1}, via multi-regge theory, as a ``wee parton component'' of scattering reggeon states. As a requirement of reggeon unitarity, therefore, the supercritical pomeron is necessarily linked to ``supercritical states''. The resulting supercritical expansion produces\cite{arw1,arw5}
\begin{itemize}
\item{\it A supercritical regge pole pomeron with $\alpha_{\spom} (0)< 1$, together 
with an {\bf exchange degenerate vector regge pole}.}
\item{\it Singular, ``vector exchange'' interactions due to the ``wee parton condensate'' in states.}
\end{itemize}
The presence of a single vector reggeon immediately suggests a broken gauge symmetry, with color superconducting QCD (SU(3) color broken to SU(2)) as the obvious candidate!! However, an asymptotically free scalar can provide a smooth connection between  superconducting and unbroken QCD only if asymptotic freedom is saturated. This is realistic if, and only, if color sextet quarks are involved and                                                                                                                                                                                                       produce electroweak symmetry breaking\cite{wm} via sextet pions. Moreover, as I will come to, non-trivial properties of the pomeron condensate require massless fermion anomalies.
 
\subhead{2.  Multi-Regge Gauge Theory}

\vspace{0.1in}
\noindent {\bf 2-to-N} {\it - some brief comments}

Since the ground-breaking paper of Fadin, Kuraev, and Lipatov\cite{fkl}, the main
focus of multi-regge theory has been on physical, {\it single rapidity axis,} 2-to-N production processes. The techniques applied have become increasingly sophisticated and effective actions have been developed for both QCD and gravity.
Physical applicability for QCD requires, of course, going well beyond standard perturbation theory and a major concern 
is whether this can be consistent with confinement.

The physical graviton complicates gravity. Indeed, there can not be a gravity S-Matrix with standard analyticity properties, since reggeon unitarity would imply that multi-graviton reggeon exchange destroys polynomial boundedness. A simple resolution is that gravity is not quantized! This is probably also necessary if particle physics is described by a bound-state S-Matrix without off-shell amplitudes, as I will advocate. 

\vspace{0.1in}
{\it Although the multiple rapidity axis kinematics that I discuss next may appear, at first sight, to be unphysical compared with
single rapidity multi-regge theory, what follows may, ultimately, be understood as
far more physical, and could, perhaps, be the future of multi-regge theory? We consider multiple rapidity axes involving large orthogonal space-like momenta.}

\vspace{0.1in}
\noindent {\bf Non-Planar N-to-N}

The key dynamical elements are fermion triangle anomaly vertices\cite{arwan} and reggeon kernels.
As illustrated in Fig.~2, with three rapidity axes {\it ($\perp$ in space)}, $\gamma$- matrices accumulate in non-planar
fermion loop interactions of multiple gauge reggeons to form anomaly generating
(reggeon vertex) triangle diagrams.
\begin{center}
\epsfxsize=5.3in\epsffile{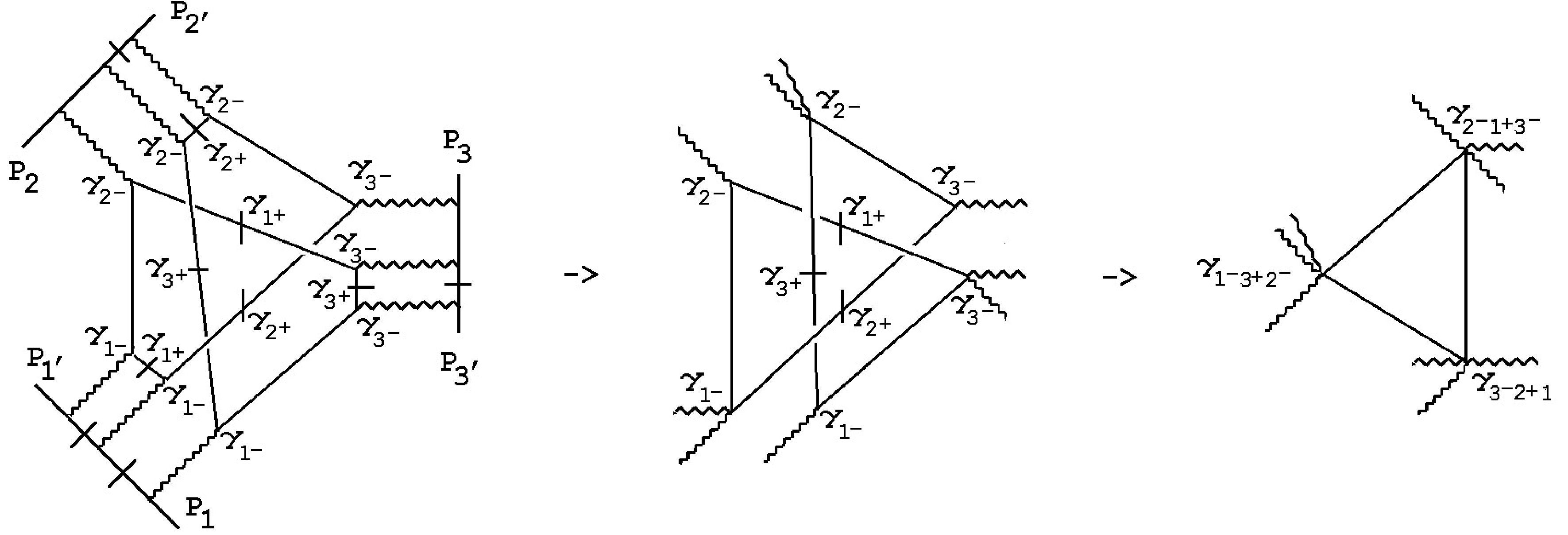}

Figure 2. An ``Anomaly Vertex'' Triangle Diagram Reggeon Interaction
\end{center}

If the fermions are massless, then anomaly vertices contain a $q^2=0$ anomaly pole
generated by a zero momentum chirality transition\cite{arw10}. The generation of a pion anomaly pole  
is illustrated in Fig.~3 
\begin{center}
\epsfxsize=5in\epsffile{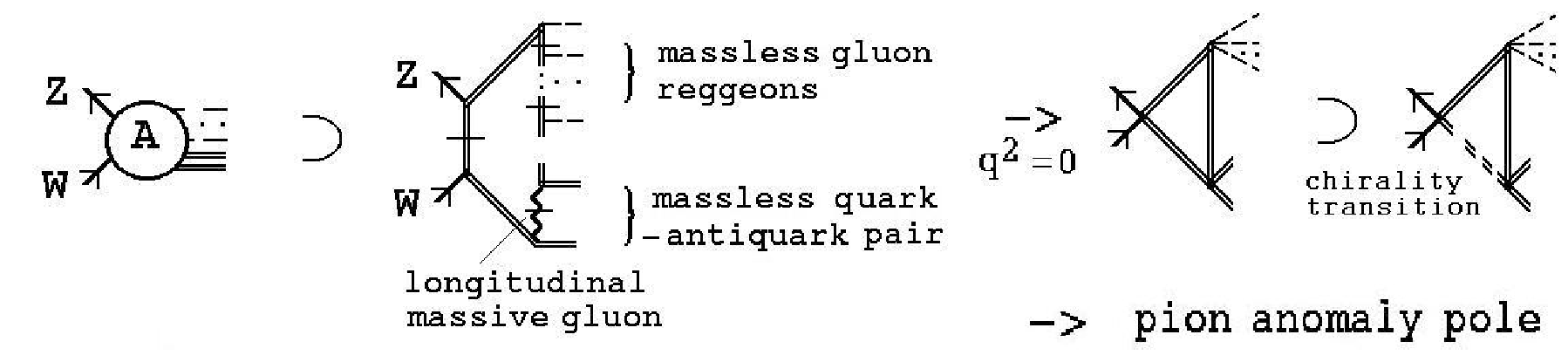}

Figure~3. Generation of a Pion Anomaly Pole
\end{center}

Gauge boson reggeon kernels determine the infra-red 
divergences coupling to anomaly poles. Most familiar are massless gluon kernels, illustrated in Fig.~4, that are generalizations of the BFKL kernel 
\begin{center}
\parbox{1.8in}{
\epsfxsize=1.5in 
\epsffile{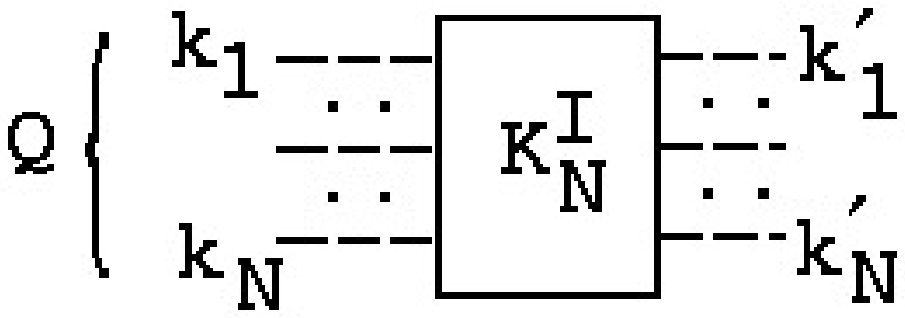}}
\parbox{1.7in}{
$$
=~K^I_N(\underline{k},\underline{k}')~\to ~\infty~,~~~Q^2, I \neq ~0
$$}

$~$
\newline $~$
\newline 
Figure 4. Massless Gluon Kernels

\end{center}
Also important are massless gluon reggeon interactions with other reggeon states, as illustrated in Fig.~5.
\begin{center}
\epsfxsize=5in
\epsffile{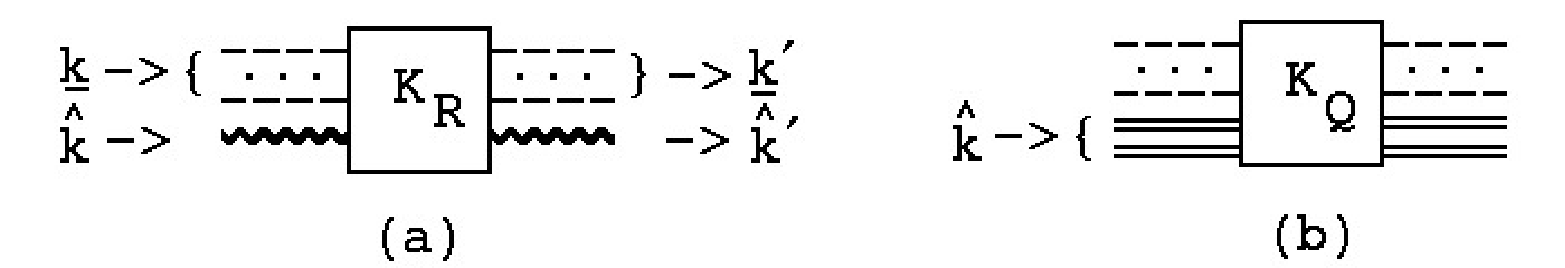}

Figure~5. Massless gluon reggeon interactions with (a) a massive reggeized gluon
$~$ (b) a quark-antiquark pair.
\end{center}
Crucially, as illustrated in Fig.~6, for anomalous \{$c\neq \tau$\} massless multi-gluon
combinations \footnote{C.f. the anomaly current $\sim$ anomalous three gluon combination. Anomalous gluon reggeon states can be thought of as ``non-local'' generalizations of the anomaly current.} there are no reggeon interaction kernels.
\begin{center}
\epsfxsize=6in
\epsffile{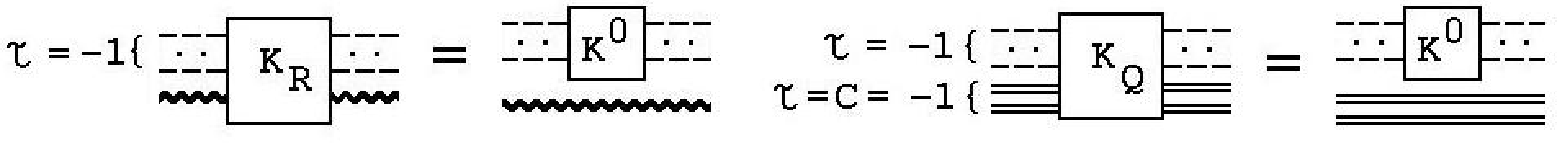}

Figure~6. Anomalous gluons have no interaction kernels.
\end{center}

In superconducting QCD, anomalous gluons couple only to anomaly poles and so, having no interaction kernels, they survive the exponentiation
of reggeization infra-red divergences. Consequently, an overall (``wee parton'')
gluon infra-red scaling divergence
generates a ``pomeron condensate'' - in both the pomeron and reggeon states. 
The supercritical pomeron then appears directly\cite{arw5} in di-triple-regge reggeon diagrams of the form shown in Fig.~7.

\vspace{0.1in}
\begin{center}
$~~$ \epsfxsize=3in
\epsffile{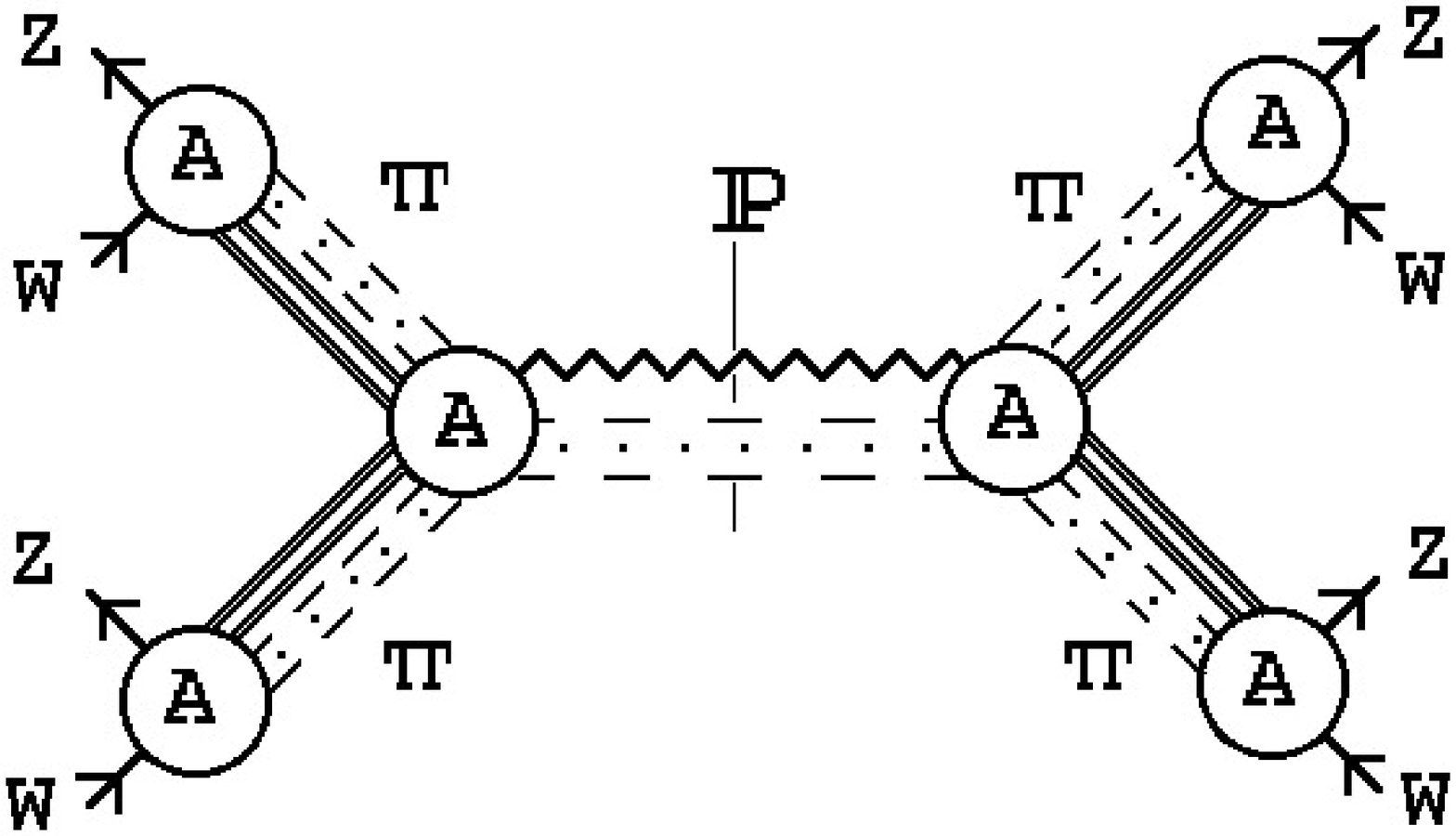}
\hspace{0.3in}\epsfxsize=2.5in
\epsffile{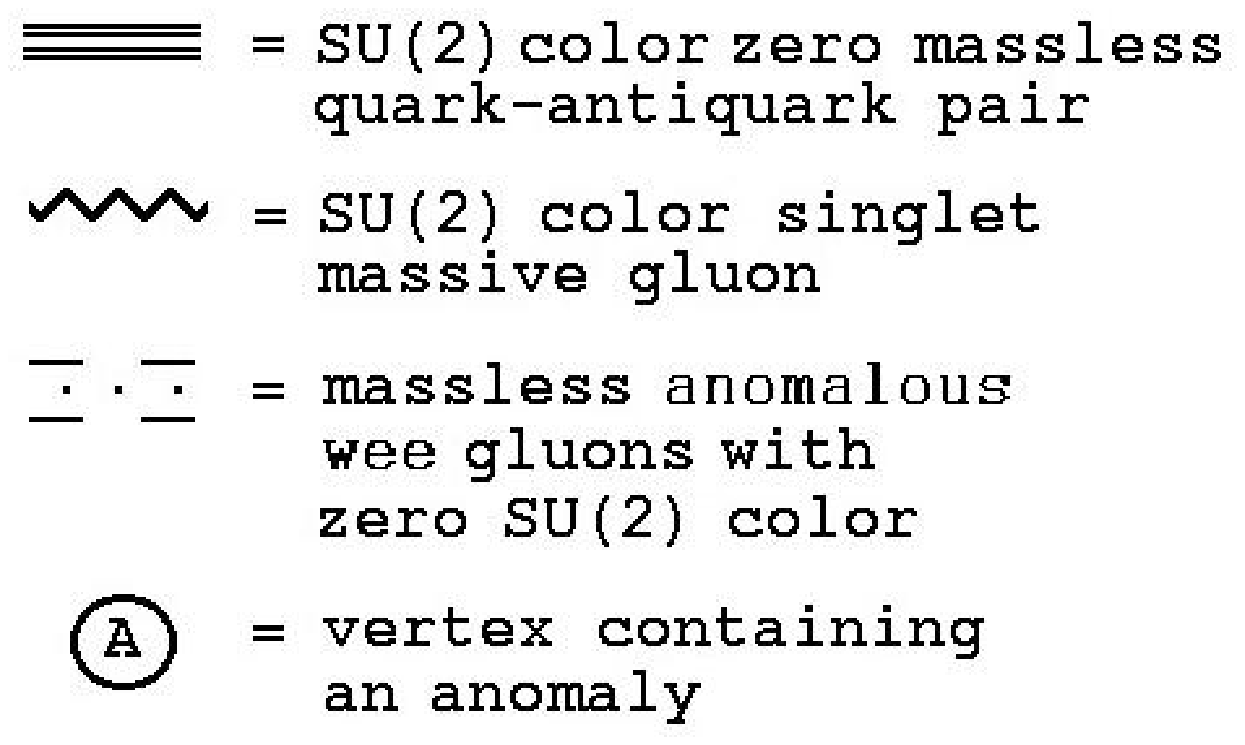}

$~$

Figure~7. Di-triple reggeon diagrams containing the supercritical pomeron 
\end{center}  

$SU(2)$ anomalous gluons have $\tau=-1 = -C$ and so
the pomeron is a $\tau=+1$ regge pole which is
exchange degenerate with a reggeized massive gluon, just as in 
supercritical RFT. The anomalous, $C= -1$, color charge parity of the pomeron is directly linked to the chiral symmetry breaking nature of the anomaly pole bound-states. Anomaly pole ``pions'' contain an unphysical zero momentum quark (or anti-quark) that undergoes a chirality transition and becomes physical via the production of anomalous wee gluons. 

The final result is a confining, chiral symmetry breaking, spectrum generated in superconducting QCD.

\subhead{4. The Critical Pomeron in Massless QCD$_S$}

\vspace{0.1in}
SU(3) color gives\cite{arw5,arw10} the Critical Pomeron, provided the wee gluon condensate survives via an infra-red fixed-point. This requires  massless QCD$_S$, i.e. 6 triplet quarks and 2 sextet quarks, which are all massless. The following list is amongst the experimentally desirable features that appear\cite{arw05}.

\begin{itemize} 
\item{\it Bound-states are triplet and sextet (pseudoscalar) mesons, together with 
(+ quark reggeon) triplet and sextet baryons. ``Neusons'', i.e. sextet neutrons, can provide DARK MATTER.}
\item{{\it There are} NO {\it hybrid sextet/triplet states.}} 
\item{{\it There are}  NO {\it glueballs,} NO {\it BFKL pomeron, and} NO {\it odderon.}
The BFKL pomeron and the odderon do not couple to the chiral symmetry breaking, anomaly pole generated,
bound-states.}
\item{\it The Critical Pomeron is a factorized regge pole, plus a triple pomeron interaction, that should give parton model factorization.}
\item{\it Sextet anomaly color factors imply larger masses for sextet states, including
dark matter neusons..} 
\item{\it Wee gluon color factors imply larger high-energy pomeron cross-sections for sextet states, including dark matter!}
\end{itemize}

 The critical behavior involves the zero momentum quark/antiquark chirality transitions 
(Dirac sea shifts), that are due to initial SU(2) anomalous wee gluons, becoming random dynamical fluctuations associated with general anomalous wee gluons  within the full SU(3) group\footnote{Since anomalous wee gluons will couple to instantons, it is conceivable that these fluctuations could be associated with topological fields, away from the light-cone kinematic regime.}. A priori, the massless quark infra-red fixed-point implies scale-invariant infra-red amplitudes that could not contain massive particle states. Consequently, if a bound-state S-Matrix with massive particles exists, there must be no off-shell correlation functions, which would necessarily have to be scale-invariant. A big problem is, however, that
massless QCD$_S$ has a huge chiral symmetry - a very large triplet quark 
symmetry and a sextet quark symmetry. As a result, the S-Matrix would have many massless Goldstones and a correspondingly serious
infra-red problem.

At this point, it would seem to be very difficult, if not impossible, for the Critical Pomeron to appear in a massive hadron S-Matrix!! Fortunately, 
the electroweak interaction solves the problem.

\subhead{5. The Electroweak Interaction plus Massless QCD$_S$:}

\vspace{0.1in}
QCD sextet quarks with the right electroweak quantum numbers, plus the electroweak interaction, embed
uniquely\cite{kw} (with asymptotic freedom and no anomaly) in 
QUD\footnote{QUD $\equiv$ Quantum Uno/Unification/Unique/Unitary/Underlying Dynamics}, i.e. SU(5) gauge theory with
left-handed massless fermions in the $5 \oplus 15 \oplus
40 \oplus 45^*$ representation.
Under $ SU(3)\otimes SU(2)\otimes
U(1)$

\openup-0.5\jot{ 
{\footnotesize $$ 
5=(3,1,-\frac{1}{3}))
+(1,2,\frac{1}{2})~,~~~~~ 15=(1,3,1)+
(3,2,\frac{1}{6}) + (6,1,-\frac{2}{3})~,
$$
$$
40=(1,2,-\frac{3}{2})
+(3,2,\frac{1}{6})+
(3^*,1,-\frac{2}{3})+(3^*,3,-\frac{2}{3}) + 
(6^*,2,\frac{1}{6})+(8,1,1)~,
$$
$$
45^*=(1,2,-\frac{1}{2})+(3^*,1,\frac{1}{3})
+(3^*,3,\frac{1}{3})+(3,1,-\frac{4}{3})+(3,2,\frac{7}{6}))+
 (6,1,\frac{1}{3}) +(8,2,-\frac{1}{2})
 $$}}
 
\vspace{0.01in}
Astonishingly, there are 3 ``generations'' of both leptons and triplet quarks.
Only the sextet quarks are input. The octet leptoquarks and the triplet quarks, together with the leptons, are all added. QUD is vector-like with respect to SU(3)xU(1)$_{em}$, but the SU(2)xU(1) quantum numbers are not quite right. Fortunately, as with massless QCD$_S$, all the massless elementary fermions are 
confined via anomaly dynamics but, in this case, there are no problems for the existence of a massive S-Matrix \footnote{Note that quark masses will appear in L$_{eff}$ for the physical, bound-state, leptons.}.
 
QUD has the full fermion content needed to reproduce the 
Standard Model\cite{arw10,arw07,arw08}. It has no exact chiral symmetries, but it does have an infra-red fixed-point. The sextet quark sector produces the much desired ``Beyond the Standard Model Physics'' of electroweak symmetry breaking and dark matter. In addition, the color octets can produce the correct generation structure of the Standard Model via large $k_{\perp}$ anomaly poles. Even more remarkably, very small neutrino masses should be a direct consequence of the very small QUD coupling. 

It surely seems very likely, as I have argued for some time, that the massless fermion anomaly 
dynamics of the uniquely unitary Critical Pomeron appears in the massive bound-state QUD S-Matrix, with the interactions and low mass spectrum of the Standard Model (including massive neutrinos) uniquely
required, in effect, to be present to obtain a massive S-Matrix.
\begin{center}
{\bf The origin of the Standard Model could be 
that it is reproducing the Unique, Unitary, S-Matrix !!}
\end{center}
An outline of my arguments follows. As I will comment upon further at the end of the talk, more details are obviously needed.

\vspace{0.2in}
\noindent{\bf The Massless Limit of QUD Reggeon Diagrams}

\vspace{0.1in}
QCD$_S$ chirality transitions do not conflict with the vector gauge symmetry, while in the QUD S-Matrix they break the {\bf non-vector} part of the gauge symmetry.
The chirality transitions appear only in anomaly pole reggeon vertices and they are produced by the zero mass limit. Consequently, how masses and $\lambda_{\perp}$ are removed\cite{arw10} is crucial!\footnote{In the QUD S-Matrix the transitions will be dynamically randomized via the Critical Pomeron.}

Using 24 and 5$\oplus$5$^*$ scalars gives masses to all fermions. Using only 
5$\oplus$5$^*$ scalars for the gauge bosons gives a smooth massless limit via complementarity.
Decoupling fermion scalars first, leaves chirality 
transitions that break SU(5) to SU(3)$_C\otimes$U(1)$_{em}~$ {\bf in anomaly vertices only.} Decoupling gauge boson scalars successively gives reggeon
global symmetries
\begin{center}
$
\rightarrow ~SU(2)_C~, ~~\rightarrow~  SU(4)~~, ~~~
 \lambda_{\perp} \to \infty~,~~  \rightarrow ~~ SU(5)
$
\end{center}
It is crucial that the last scalar removed is asymptotically free and so the $\lambda_{\perp} \to \infty$ limit can be taken between the SU(4) and SU(5) limits.

As in QCD$_S$,  anomalous gauge boson wee partons appear that couple via chiral Goldstone anomaly poles. Their complexity increases\cite{arw10} with each limit.
\begin{itemize}
\item{After the SU(2)$_C$ limit, the wee gluons, bound-states, and interactions of color superconducting QCD appear.} 
\item{ In the SU(4) limit, with $\lambda_{\perp} \neq \infty $, many fermion loops violate Ward identities and the exponentiation of divergences goes well beyond reggeization. Exchanges of left-handed gauge bosons not mixed with sextet pions
are eliminated. Also left-handed gauge bosons do not contribute as wee partons.}
\item{When the SU(5) limit is taken, after the $\lambda_{\perp} \to \infty$ limit, the Critical Pomeron and the massless photon appear together. Also, color octet chiral
anomalies at $ k_{\perp} = \infty$ produce bound-states in Standard Model generations.}
\end{itemize}

The ``universal wee partons'' that we finally obtain are combinations of SU(5), vector coupling, anomalous wee gauge bosons in the adjoint representation.
Vector interactions between bound-states, appear in diagrams of the form shown in Fig.~8.

\vspace{0.2in}
\begin{center}
\epsfxsize=5.5in
\epsffile{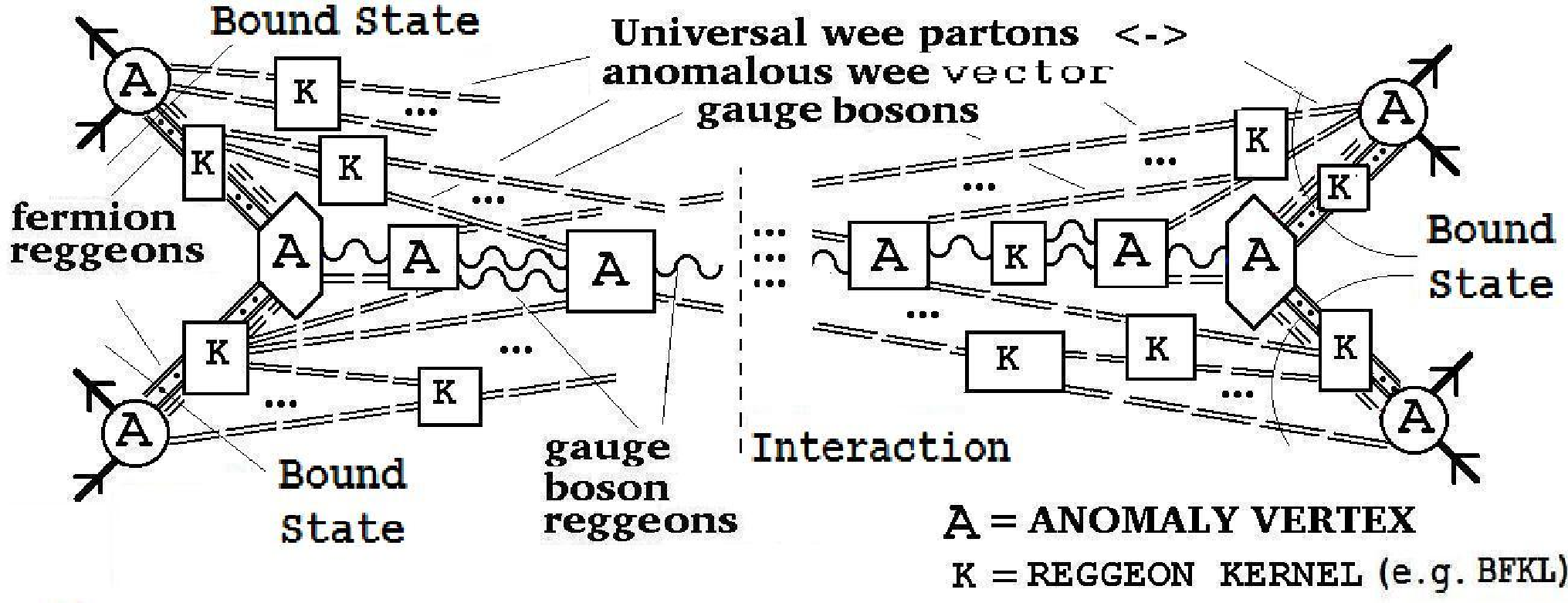}

\vspace{0.2in}
Figure~8. Di-Triple-Regge QUD reggeon diagrams containing a vector interaction
\end{center} 

 The interactions preserve vector SU(3)xU(1)$_{em}$ and couple via anomalies. They are\cite{arw10} Standard Model interactions.
\begin{enumerate}\openup1\jot{
\item{{\bf The Even Signature Critical Pomeron} is, approximately, an SU(3) gluon reggeon plus wee gauge bosons that, together, produce an SU(5) singlet projection. 
As in massless QCD$_S$, there is no BFKL pomeron and no odderon.}
\item{{\bf The Odd Signature Photon} is, approximately, a U(1)$_{em}$
gauge boson + wee gauge bosons that, together, produce an SU(5) singlet projection.}
\item{{\bf The Electroweak Interaction} is, approximately, a left-handed gauge  boson that mixes with a sextet pion (via anomalies) and carries sextet quantum numbers and with wee gauge bosons again producing an SU(5) singlet projection.}}
\end{enumerate}

Anomaly color factors in wee gauge boson infinite sums enhance
couplings - hopefully to Standard Model values, i.e. 
$$\alpha_{\scriptscriptstyle QCD} ~> ~\alpha_{\scriptscriptstyle em}~>>~ \alpha_{\scriptscriptstyle QUD} \sim
\frac{1}{120}$$
For bound-states, anomaly vertex mixing, combined with fermion and wee parton color
factors, also produces a wide range of mass scales. Very significantly, as already noted,
the very small QUD coupling should result in desirably small neutrino masses.

Standard Model generations of physical hadrons and leptons appear  
via infinite light-cone momentum octet anomalies as follows.

\vspace{0.2in}
\noindent {\bf Leptons\cite{arw10,arw11}} 

\vspace{0.1in}
Lepton bound states contain three elementary leptons -
\begin{itemize}
\openup2\jot{
\item{\it $(e^-,\nu)$ ~$ \leftrightarrow ~
(1,2,-\frac{1}{2}) \times \{(1,2,-\frac{1}{2}) (1,2,\frac{1}{2})\}_{AP} \times
\{(8,1,1)(8,2,-\frac{1}{2})\}_{UV}$ }
\item{\it $(\mu^-,\nu)$ ~ $ \leftrightarrow ~
(1,2,\frac{1}{2})\times  \{(1,2,-\frac{1}{2})  (1,2,-\frac{1}{2})\}_{AP} 
\times \{(8,1,1)(8,2,-\frac{1}{2})\}_{UV}$}
\item{\it $(\tau^-,\nu)$~  $ \leftrightarrow~
(1,2,-\frac{3}{2}) \times \{(1,2,\frac{1}{2})  (1,2,\frac{1}{2})\}_{AP} 
\times \{(8,1,1)(8,2,-\frac{1}{2})\}_{UV}$}}
\end{itemize}

\vspace{0.1in}
\noindent Anomaly color factors imply 
$$M_{\scriptstyle hadrons} >> M_{\scriptstyle leptons} >>  M_{\scriptstyle \nu 's} ~\sim~
\alpha_{\scriptstyle QUD}$$
The electron is almost elementary since, in effect, the anomaly pole disturbs the Dirac sea minimally. The muon has the same constituents, but in a dynamical configuration
that will obviously generate a significant mass.

\vspace{0.2in}
\noindent{\bf Hadrons}\cite{arw11,arw12,arw13} 

\begin{enumerate} 
\item{Two QUD triplet quark generations give Standard Model hadrons - that mix appropriately.}
\item{The physical b quark is a mixture of all three QUD generations}.
\item{Sextet pions produce electroweak symmetry breaking by mixing with
left-handed vector bosons.} 
\item{Sextet neutrons, \{``neusons''\} are stable and so will provide DARK MATTER
- with many desirable properties.}
\item{The left-handed ``top quark'' mixes with 
exotic quarks and has no low mass states.}
\item{ ``Top quark physics'' is very different from Standard Model top physics. However, because the sextet ${\eta}$  produces the Standard Model final states, it is experimentally hard to distinguish the difference.}
\item{Mixing of triplet and sextet $\eta$ states gives two mixed-parity scalars, the
$\eta_3$ and the $\eta_6$.}
\item{The mass of the $\eta_3$ could  be $\sim$ 125 GeV  and so it could be a candidate for the ``Higgs boson'' discovered at the LHC. It is primarily a ``$t\bar{t}$'' resonance, where the $t$ is the QUD top quark!} 
\item{The $\eta_6$ should be seen (and perhaps has been seen) at the LHC - at the
$t\bar{t}$ threshold, where $t$ is the Standard Model top quark.}
\item{``Tree-unitarity'' suggests the combined     
$\eta_3$ and $\eta_6$ couplings should reproduce the Standard Model 
electroweak couplings of the Higgs boson.}
\end{enumerate}

\vspace{0.2in}
 \noindent {\bf Existing (interpreted) evidence} for the QCD sextet quark sector includes\cite{arw11,arw12,arw13,arw132}: 
   
 \noindent \begin{enumerate} 
 \item{``Heavy Ion'' UHE cosmic rays are dark matter neusons.} 
 \item{The spectrum knee is due to arriving/produced neuson
 thresholds.} 
 \item{Enhancement of high multiplicities and small $ p_{\perp}$
 at the LHC reflects a sextet anomaly generated triple pomeron coupling.}
 \item{``Top quark events" are due to the $\eta_6$ 
 resonance - interference with the background produces the Tevatron asymmetry.}
 \item{$W^+W^-$ and $Z^0Z^0$ pairs have high mass excess x-sections, with the
  $\eta_6$ resonance appearing at the ``$t\bar{t}$ threshold".}
 \item{The 125 GeV Higgs is the QUD \{$t_R\bar{t}_L $ + $\eta_6$\}
 resonance.}
 \item{The AMS $e^+/e^-$ ratio reflects EW scale CR production of W's \&
 Z's  (+ neuson/antineuson annihilation?)}
 \item{Low luminosity Tevatron/LHC events with a $Z$ pair + high
 multiplicity 
 of small  $p_{\perp}$ particles, could be QUD  (not Standard Model) events. }
 \item{TOTEM+CMS missing momentum events could be $~ZZ ~\to \nu$'s }
 \end{enumerate}

\newpage 

To illustrate the complexity of the underlying wee parton physics in bound-state 
scattering, we show the simplest QUD amplitude, i.e.
electron scattering via photon exchange as illustrated in Fig.~9. 

\vspace{0.2in}
\begin{center}
\epsfxsize=2.9in
\epsffile{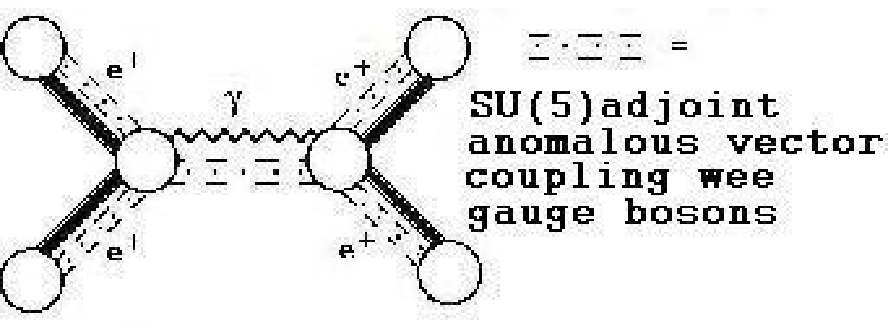}$~~$
\epsfxsize=2.9in
\epsffile{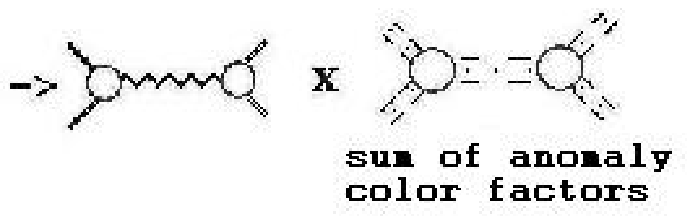}

\vspace{0.4in}

\epsfxsize=6.2in
\epsffile{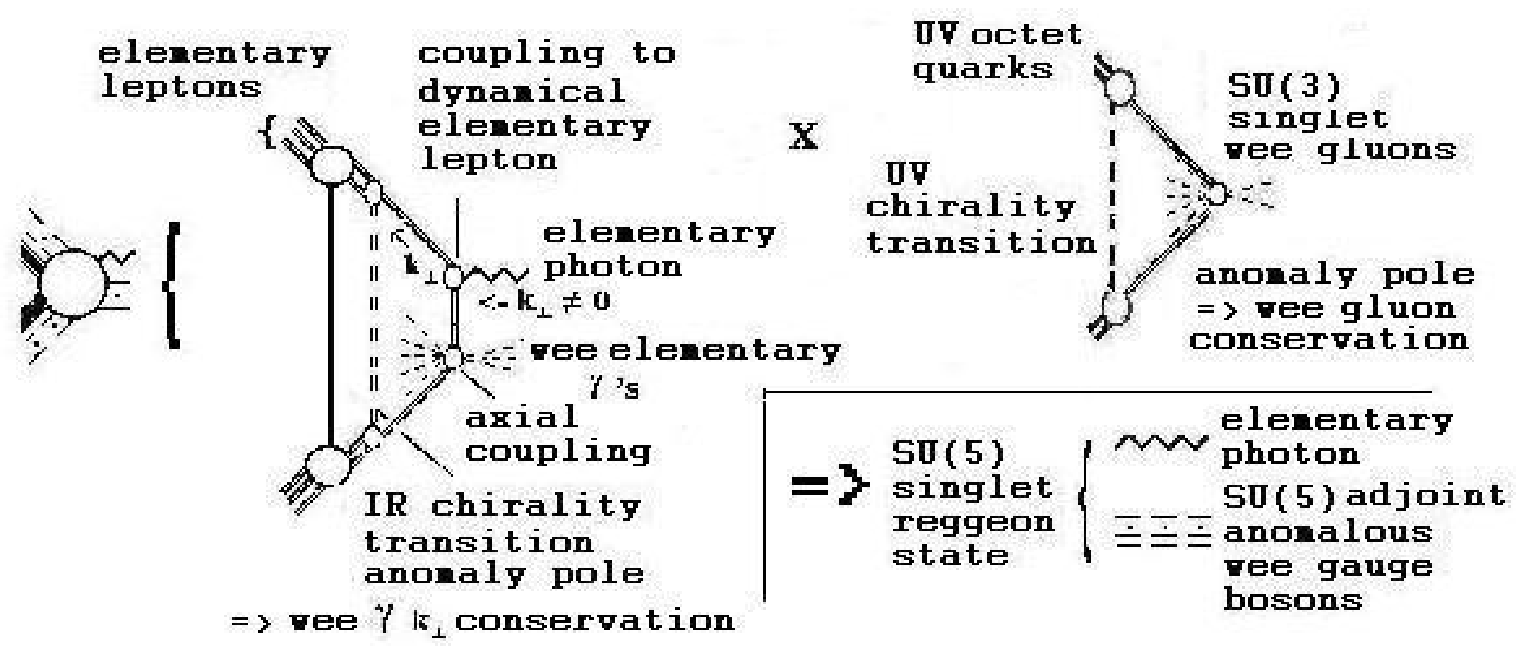}

\vspace{0.4in}
\epsfxsize=6.2in
\epsffile{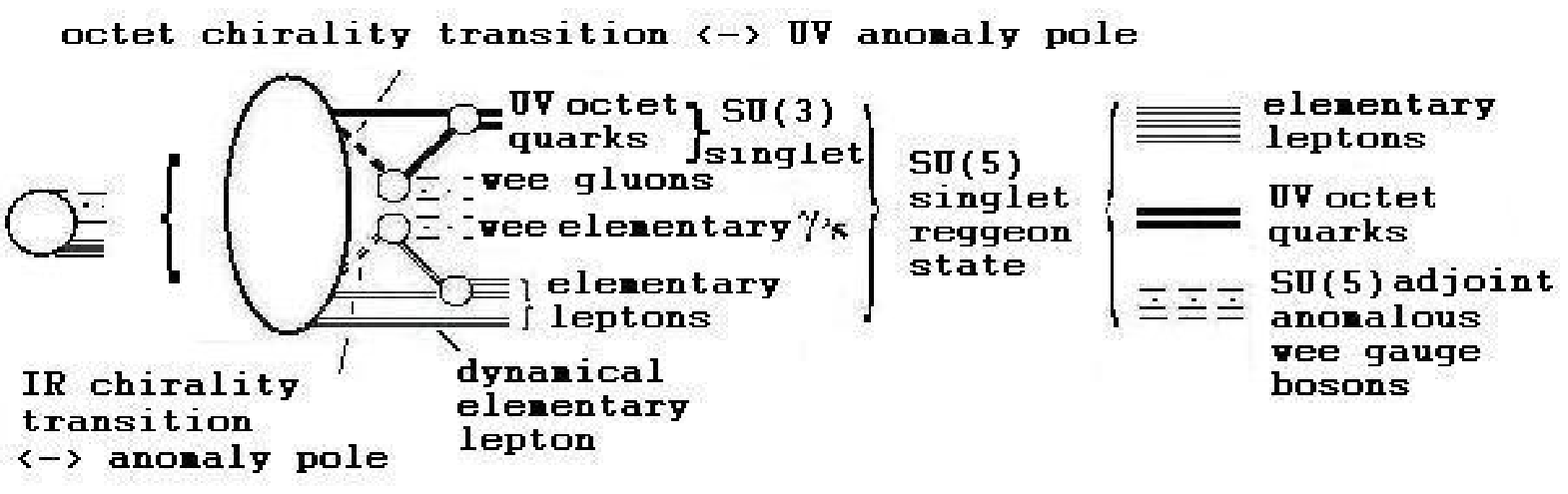}

\vspace{0.3in}
Figure~9. Electron scattering via photon exchange

\end{center}

\newpage 

\subhead{Some QUD Virtues}

\vspace{0.2in}
\begin{itemize} 
\item{QUD is self-contained and is either entirely
right, or simply wrong!}
\item{ The scientific and aesthetic importance of an 
underlying and unifying (non-SUSY) massless field theory for the Standard Model 
can not be exaggerated.}
\end{itemize}

\vspace{0.1in}
\noindent Assuming the QUD S-Matrix can be derived as I have outlined, then -
 \begin{enumerate} 
\item{The only new physics is a high mass sector of the strong
interaction that gives electroweak symmetry breaking and dark matter.}
\item{Parity properties of all the interactions are explained by the anomaly dynamics.}
\item{ Confinement, chiral symmetry breaking, the parton model, and the Critical Pomeron all appear in QCD - in a form more consistent with experiment 
than conventional expectations.}
\item{The massless photon partners the ``massless'' Critical Pomeron.}
\item{Anomaly vertex mixing implies wee parton color
factors produce a wide range of (Standard Model ???)
scales and masses, with small neutrino masses due
to the very small QUD coupling.}
\item{Particles and fields are truly distinct. Hadrons and leptons have equal status.}
\item{Symmetries and masses are dynamical S-Matrix
properties. There are no off-shell amplitudes and there is no Higgs field.}
\item{As a massless, asymptotically free,  
fixed-point theory, with no renormalon-related vacuum condensates, QUD induces
Einstein gravity with zero cosmological constant.}\end{enumerate}


\subhead{6. The Nightmare Scenario:}

\vspace{0.2in}
In his overview at Strings 2013, David Gross discussed the ``extreme pessimistic scenario'' in which the Standard Model Higgs boson is discovered at the LHC but no other new (short-distance) physics, in particular no signal for SUSY, is seen.

He said publicly, what many others have said privately, that this scenario was looking more and more likely and (if it is established) then, he acknowledged,
\begin{center}
 {\bf ``We Got it Wrong. How did we misread the signals? What to Do?''}
\end{center}
He said that the field, and string theorists in particular, will suffer badly. 
Younger theorists must figure out where previous generations went wrong and suggest what experimenters should now look for.

\newpage If supersymmetry and, by inference, string theory are not invoked by nature in the extension/unification of the Standard Model, a huge proportion of theoretical high-energy physics research over the last few decades will have been critically wasted.
Vast resources will also have been wasted in experimental searches.

If hard evidence of an electroweak scale strong interaction
appears, supporting the existence of QUD, 
there will be a (badly needed ???) radical redirection of the field. 
Away from the (misguided ???) pursuit of ``rare'', non-existent, short-distance physics to full-scale study of novel high-energy, large x-section, long-distance physics.



\vspace{0.2in}
 \subhead{7. The Future} 
  
\vspace{0.2in}
Going forward  by formulating new short-distance theories, supersymmetric or otherwise, while ignoring all infra-red problems of QCD and all previously obtained long-distance experimental knowledge, may ultimately be seen as a major source of the nightmare scenario\footnote{In a forthcoming paper - {\it The Nightmare Scenario and the Origin of the Standard Model.
``We Got it Wrong ...How did we misread the signals? ... What to Do?''} - I will enlarge on the previous subsection from this perspective.}.

If I am right, a bound-state unitary S-matrix, with all the pomeron
forward physics properties seen in experiments, is a very special property of a unique massless field theory in which the infra-red anomaly dynamics is far, far, from the, short-distance based, current theory paradigm. Multi-regge theory, alone I believe, offers the possibility to derive the bound-state amplitudes that would be needed to study QUD physics at the LHC.

As is surely evident, I have pushed my application of multi-regge theory far beyond existing, established, results. Very extensive development is needed
to provide a practical framework. But, if there is serious experimental encouragement, the development could be fast and furious. A short, but far from inclusive, list of high
priority topics might be as follows.

\vspace{0.1in}
\begin{enumerate}
\item{Detailed extraction of bound-state amplitudes from multi-regge/helicity pole limits.} 
\item{Characterization of limits in which anomaly vertices appear.}
\item{Derivation and classification, with explicit expressions, of anomaly vertices.}
\item{Summation, via integral formulae(?), of wee parton enhancement of elementary couplings.}
\item{Comprehensive study of anomalous wee gluon kernels.}
\item{Full emergence of the Standard Model via the removal of masses and 
$\lambda_{\perp}$ in QUD  reggeon diagrams.}
\end{enumerate}

\end{document}